\documentclass[prl,twocolumn,superscriptaddress]{revtex4-2}

\usepackage{amssymb,amsmath,bm,natbib}
\usepackage{color}
\usepackage{slashed}
\usepackage{graphics}
\usepackage{graphicx}
\usepackage[utf8]{inputenc}
\usepackage{upgreek}
\usepackage[caption=false]{subfig}
\usepackage{hyperref}
\usepackage{url}
\usepackage{dsfont}
\usepackage{float} 
\usepackage{cancel}
\usepackage{units}
\usepackage{blindtext}
\usepackage[dvips]{feynmp}
\usepackage{upgreek}
\renewcommand{\eqref}[1]{\mbox{Eq.~(\ref{#1})}}

\newcommand{\figref}[1]{\mbox{Fig.~\ref{#1}}}

\begin{document}
\preprint{}
	\title{New Constraint for Isotropic Lorentz Violation from LHC Data}

\author{David Amram}
\email{d.amram@ip2i.in2p3.fr}
\affiliation{Universit\'{e} de Lyon, Universit\'{e} Claude Bernard Lyon 1, CNRS/IN2P3, IP2I Lyon, UMR 5822, Villeurbanne, France}

\author{Killian Bouzoud}
\email{kbouzoud@subatech.in2p3.fr}
\affiliation{SUBATECH, Nantes Universit\'e, IMT Atlantique, IN2P3/CNRS, 4 rue Alfred Kastler, La Chantrerie BP 20722, 44307 Nantes, France}

\author{Nicolas Chanon}
\email{nicolas.pierre.chanon@cern.ch}
\affiliation{Universit\'{e} de Lyon, Universit\'{e} Claude Bernard Lyon 1, CNRS/IN2P3, IP2I Lyon, UMR 5822, Villeurbanne, France}

\author{Hubert Hansen}
\email{hansen@ipnl.in2p3.fr}
\affiliation{Universit\'{e} de Lyon, Universit\'{e} Claude Bernard Lyon 1, CNRS/IN2P3, IP2I Lyon, UMR 5822, Villeurbanne, France}

\author{Marcos R. Ribeiro Jr.}
\email{marcosribeiro@usp.br}
\affiliation{Instituto de F\'{i}sica, Universidade de S\~{a}o Paulo, Cidade Universit\'{a}ria \\ S\~{a}o Paulo (SP), 05508-090, Brazil}

\author{Marco Schreck}
\email{marco.schreck@ufma.br}
\affiliation{Departamento de F\'{i}sica, Universidade Federal do Maranh\~{a}o \\
Campus Universit\'{a}rio do Bacanga, S\~{a}o Lu\'{i}s (MA), 65085-580, Brazil}

\begin{abstract}
New calculations for the kinematics of photon decay to fermions \textit{in vacuo} under an isotropic violation of Lorentz invariance (LV), parameterized by the Standard-Model Extension (SME), are presented in this paper and used to interpret prompt photon production in LHC data. The measurement of inclusive prompt photon production at the LHC Run 2, with photons observed up to a transverse energy of $\unit[2.5]{TeV}$, provides the lower bound $\Tilde{\kappa}_{\mathrm{tr}} > -1.06 \times 10^{-13}$ on the isotropic coefficient $\Tilde{\kappa}_{\mathrm{tr}}$ at 95\% confidence level. This result improves over the previous bound from hadron colliders by a factor of 55. The calculations for the kinematics of photon decay have further potential use to constrain LV coefficients from the appearance of fermion pairs, for instance, top-antitop.
\end{abstract}



\maketitle

\section{Introduction}

Considering the technology and experimental capabilities of the early 21st century, it is not possible to probe quantum gravity directly, since experiments would require particles whose energies lie in the vicinity of the Planck scale. Only at such an energy one would be able to resolve structures of the smallest length scales imaginable such as strings or a spacetime foam \cite{Wheeler:1957mu,Hawking:1978pog}. Fortunately, there are alternatives to observing quantum gravity phenomena in a direct way. We resort to the two most successful theories: the Standard Model (SM) of elementary particles and General Relativity (GR). The first is based on global Lorentz invariance, whereby GR is governed by diffeomorphism symmetry as well as local Lorentz invariance. It is these symmetries that play a crucial role in quantum gravity.

The SM and GR describe particles and gravity astoundingly well for energies much lower than the Planck scale. A valuable approach to quantum gravity phenomenology consists in searching for tiny deviations of these established theories. In the framework of effective field theory, there are several possibilities for how to proceed. First, the SM is extended by excitations of new degrees of freedom described by non-SM fields. Second, contributions are introduced that preserve the symmetries of the SM and GR, but involve higher-order derivatives. Third, an alternative is to consider violations of at least one of the fundamental spacetime symmetries. Our interest lies on the latter approach.

The gravitational field can be neglected for our purpose, which means that we will focus on violations of global Lorentz invariance. A motivation is provided by a set of articles indicating a breakdown of Lorentz symmetry in string field theory \cite{Kostelecky:1988zi,Kostelecky:1989jp,Kostelecky:1989jw,Kostelecky:1991ak,Kostelecky:1994rn}, loop quantum gravity \cite{Gambini:1998it,Bojowald:2004bb}, noncommutative field theory \cite{Amelino-Camelia:1999jfz,Carroll:2001ws}, spacetime foam models \cite{Klinkhamer:2003ec,Bernadotte:2006ya,Hossenfelder:2014hha}, chiral field theories defined in spacetimes with nontrivial topologies \cite{Klinkhamer:1998fa,Klinkhamer:1999zh,Ghosh:2017iat}, and Ho\v{r}ava-Lifshitz gravity \cite{Horava:2009uw}. A comprehensive technique employed in a broad program of experimental searches for Lorentz violation (LV) is to employ a model-independent effective parametrization of deviations from Lorentz symmetry, which the Standard-Model Extension (SME) \cite{Colladay:1996iz,Colladay:1998fq} provides. The latter respects the gauge symmetry and particle content of the SM. Lorentz violation is parameterized by tensor-valued background fields whose components, which are denoted as controlling coefficients, describe the amount of symmetry violation. These background fields are coupled to the SM fields such that the resulting vacuum expectation values are constants with respect to coordinate boosts and rotations. Within effective field theory, {\em CPT}-violation was shown to imply a violation of Lorentz symmetry \cite{Greenberg:2002uu}, which is why {\em CPT}-violating operators are contained in the SME.

In the SME, particles obey modified dispersion relations and processes are described perturbatively by a set of modified Feynman rules \cite{Kostelecky:2001jc,Colladay:2006rk,Ferrero:2011yu}. The modified kinematics may enable unusual processes that are forbidden in the SM due to energy-momentum conservation. For example, under certain conditions, a fermion subject to LV can lose energy by the emission of photons. This process is reminiscent of Cherenkov radiation in optical media with the crucial difference that it occurs \textit{in vacuo}. Therefore, the latter is called vacuum Cherenkov radiation \cite{Beall:1970rw,Coleman:1997xq,Moore:2001bv,Lehnert:2004be,Lehnert:2004hq,Kaufhold:2005vj,Kaufhold:2007qd,Altschul:2006zz,Altschul:2007kr,Hohensee:2008xz,Klinkhamer:2008ky,Altschul:2014bba,Schober:2015rya,Diaz:2015hxa,Kostelecky:2015dpa,Colladay:2016rmy,Altschul:2016ces,Colladay:2016rsf,Colladay:2017auq,Schreck:2017isa,Altschul:2017xzx,Schreck:2017egi,Schreck:2018qlz}. Another prominent process is photon decay into a fermion-antifermion pair \cite{Liberati:2001cr,Jacobson:2001tu,Jacobson:2002hd,Jacobson:2005bg,Klinkhamer:2008ky,Hohensee:2008xz,Shao:2010wk,Altschul:2010nf,Rubtsov:2012kb,Satunin:2013an,Rubtsov:2013wwa,Diaz:2015hxa,Kalaydzhyan:2016lfo,Martinez-Huerta:2016odc,Martinez-Huerta:2017ulw,Martinez-Huerta:2017ntv,Martinez-Huerta:2017unu,Klinkhamer:2017puj,Martinez-Huerta:2019rkx,Satunin:2019gsl,Chen:2021hen,Duenkel:2021gkq,Wei:2021ite}, which is the focus of this paper.

Photon decay can occur in the {\em CPT}-even photon sector of the SME that is governed by 19 independent controlling coefficients \cite{Colladay:1998fq,Kostelecky:2002hh,Bailey:2004na}. Ten of these imply vacuum birefringence and  are strongly constrained by spectropolarimetry measurements \cite{Kostelecky:2001mb,Kostelecky:2002hh,Kostelecky:2009zp,Kislat:2018rsi,Friedman:2018ubt,Friedman:2020bxa,Gerasimov:2021chj}. The remaining 9 coefficients are nonbirefringent at first order in LV. A subset of 8 leads to anisotropic photon propagation and a single one describes Lorentz-violating effects that are spatially isotropic. Searches for anisotropic effects via small-scale laboratory experiments have provided a large number of tight bounds on such coefficients \cite{Eisele:2009zz,Herrmann:2009zzb,Hohensee:2010an,Zhang:2021sbx}, whereas astroparticle physics has led to competitive constraints on the isotropic coefficient \cite{Klinkhamer:2008ky,Diaz:2015hxa,Duenkel:2021gkq,Duenkel:2023nlk} (cf.~Tab~D16 in Ref.~\cite{Kostelecky:2008ts}).

When it comes to particle energy and propagation distance, astroparticle physics experiments have an advantage over collider experiments in constraining LV.
The highest energies of astrophysical photons detected lie in the ballpark $E\approx \unit[250]{TeV}$~\cite{Satunin:2019gsl}.
However, the drawback of astroparticle experiments is that the interpretation relies on e.g., models for the distribution of the sources and for the injection of nuclei through the source environment~\cite{Unger:2015laa}, which are necessary to describe the energy spectrum, composition, and arrival direction of astroparticles~\cite{Martinez-Huerta:2019rkx}.
In addition, information related to the intrinsic spectrum of the source, its evolution, and emission mechanism is not necessarily available~\cite{Chen:2021hen,Ellis:2005sjy,Chang:2012gq}.
By contrast, Earth-based laboratory experiments provide a reproducible, controlled source of prompt photons, well-understood in term of the SM predictions, leading to limits on LV that can be considered as less model-dependent and, hence, more conservative.
Such collider bounds are complementary to astrophysical constraints.
In this work we will demonstrate how prompt photons measured at the LHC are capable of improving the constraint on isotropic {\em CPT}-even LV by a factor larger than 50, in comparison with the present limit derived at the Tevatron.


\section{Minimal-SME photon sector}
\label{sec:sme-photon-sector}

We consider a modified quantum electrodynamics with LV in the photon sector, but standard Dirac fermions:
\begin{align}
\label{eq:action-modified-photons}
S&=\int \mathrm{d}^4x\,\bigg[-\frac{1}{4}(\eta^{\mu\varrho}\eta^{\nu\sigma}+\kappa^{\mu\nu\varrho\sigma})F_{\mu\nu}(x)F_{\varrho\sigma}(x) \notag \\
&\phantom{{}={}}\hspace{1.2cm}+\frac{1}{2}\overline{\psi}(x)\left(\gamma^{\mu}\mathrm{i}D_{\mu}-m_f\right)\psi(x)+\text{H.c.}\bigg]\,,
\end{align}
with the electromagnetic field strength tensor $F_{\mu\nu}=\partial_{\mu}A_{\nu}-\partial_{\nu}A_{\mu}$ of the \textit{U}(1) gauge field $A_{\mu}$. Furthermore, $\psi$ is a Dirac spinor field and $\overline{\psi}\equiv \psi^{\dagger}\gamma^0$ the corresponding Dirac conjugated field. All fields are defined in Minkowski spacetime with metric $\eta_{\mu\nu}$ of signature $(+,-,-,-)$. We employ the standard Dirac matrices satisfying the Clifford algebra $\{\gamma^{\mu},\gamma^{\nu}\}=2\eta^{\mu\nu}$ and $m_f$ is the fermion mass. The Dirac field is minimally coupled to the gauge field via the covariant derivative $D_{\mu}=\partial_{\mu}+\mathrm{i}qA_{\mu}$ with the fermion charge $q$. 
The violation of Lorentz invariance is parameterized via $\kappa^{\mu\nu\varrho\sigma}$, which transforms as a four-tensor of rank 4 under coordinate boosts and rotations, but remains fixed with respect to boosts and rotations of experiments proper. The object $\kappa^{\mu\nu\varrho\sigma}$ implies preferred directions in spacetime leading to a boost- and/or direction-dependent form of the laws of physics, which corresponds to a violation of Lorentz symmetry.

The physics of \eqref{eq:action-modified-photons} is described by a nontrivial refractive index of the vacuum that leads to a speed of light different from the maximum velocity of Dirac particles. The refractive index may be anisotropic or even polarization-dependent resulting in vacuum birefringence. Since the latter effect is tightly constrained at the level of $10^{-34}$ to $10^{-35}$ by spectropolarimetry measurements~\cite{Kostelecky:2008ts}, we will discard the birefringent part of $\kappa^{\mu\nu\varrho\sigma}$ involving 10 coefficients. The remaining 9 coefficients are parameterized by the nonbirefringent \textit{ansatz}~\cite{Altschul:2006zz}:
\begin{equation}
\kappa^{\mu\nu\varrho\sigma}=\frac{1}{2}(\eta^{\mu\varrho}\tilde{\kappa}^{\nu\sigma}-\eta^{\mu\sigma}\tilde{\kappa}^{\nu\varrho}+\eta^{\nu\sigma}\tilde{\kappa}^{\mu\varrho}-\eta^{\nu\varrho}\tilde{\kappa}^{\mu\sigma})\,,
\end{equation}
where $\tilde{\kappa}^{\mu\nu}$ is a symmetric and traceless $(4\times 4)$ matrix. We study isotropic LV in the laboratory frame, which is the simplest choice for a first analysis before investigating more intricate cases. There is a single coefficient usually denoted as $\tilde{\kappa}_{\mathrm{tr}}$ that parameterizes isotropic LV in the theory of \eqref{eq:action-modified-photons}. The coefficients of the matrix $\tilde{\kappa}^{\mu\nu}$ are then chosen as
\begin{equation}
\tilde{\kappa}^{\mu\nu}=\frac{3}{2}\tilde{\kappa}_{\mathrm{tr}}\,\mathrm{diag}\left(1,\frac{1}{3},\frac{1}{3},\frac{1}{3}\right)^{\mu\nu}\,.
\end{equation}
Then, the dispersion relation for photons is modified and reads
\begin{equation}
\omega=\mathcal{A} \, |\mathbf{k}|\,,\quad \mathcal{A}=\sqrt{\frac{1-\tilde{\kappa}_{\mathrm{tr}}}{1+\tilde{\kappa}_{\mathrm{tr}}}}\,,
\end{equation}
where $\mathbf{k}$ is the 3-momentum of the photon. The isotropy of the result is evident. Since the phase and group velocities of photons obey
\begin{equation}
v_{\mathrm{ph}}\equiv \frac{\omega}{|\mathbf{k}|}=\mathcal{A}\,,\quad v_{\mathrm{gr}}\equiv |\boldsymbol{\nabla}_{\mathbf{k}}\omega|=\mathcal{A}\,,
\end{equation}
there is no vacuum dispersion. For $\tilde{\kappa}_{\mathrm{tr}}\in (0,1]$ we have that $v_{\mathrm{gr}}<1$. The dispersion relation of Dirac fermions is the standard one for massive particles. Hence, massive fermions can possibly propagate faster than light. For $\tilde{\kappa}_{\mathrm{tr}}\in (-1,0)$ we observe that $v_{\mathrm{gr}}>1$ and photons always move faster than fermions, no matter what is their energy.

In the first kinematic regime, vacuum Cherenkov radiation can occur, whereas photon decay \textit{in vacuo} is a characteristic process for the second regime. Both processes are governed by a threshold, i.e., they occur if the energy of the initial particle exceeds a certain minimum that depends on the fermion mass and the controlling coefficient. 
Since vacuum Cherenkov radiation with LV in photons and fermions, respectively, has already been studied exhaustively \cite{Beall:1970rw,Coleman:1997xq,Moore:2001bv,Lehnert:2004be,Lehnert:2004hq,Kaufhold:2005vj,Kaufhold:2007qd,Altschul:2006zz,Altschul:2007kr,Hohensee:2008xz,Klinkhamer:2008ky,Altschul:2014bba,Schober:2015rya,Diaz:2015hxa,Kostelecky:2015dpa,Colladay:2016rmy,Altschul:2016ces,Colladay:2016rsf,Colladay:2017auq,Schreck:2017isa,Altschul:2017xzx,Schreck:2017egi,Schreck:2018qlz}, we intend to investigate photon decay. Thus, in what follows, we will take $\tilde{\kappa}_{\mathrm{tr}}\in (-1,0)$.

\section{Kinematics and dynamics of photon decay}
\label{sec:photon-decay}

Now let us look at the basics of photon decay $\upgamma\rightarrow f\overline{f}$ within the isotropic sector of \eqref{eq:action-modified-photons}. The threshold energy $E^{\mathrm{th}}$ of the incoming photon is best computed by considering the 3-momenta of all particles being collinear. Energy-momentum conservation then implies~\cite{Klinkhamer:2008ky}
\begin{equation}
\label{eq:relation-kappa-Ethresh}
E^{\mathrm{th}}=2m_f\sqrt{\frac{1-\tilde{\kappa}_{\mathrm{tr}}}{-2\tilde{\kappa}_{\mathrm{tr}}}}\,.
\end{equation}
Several observations are in order. First, for a massless fermion, the process occurs without a threshold. Second, the expression is real only for $\tilde{\kappa}_{\mathrm{tr}}<0$, whereupon photon decay is forbidden for $\tilde{\kappa}_{\mathrm{tr}}>0$. Third, the threshold goes to infinity for $\tilde{\kappa}_{\mathrm{tr}}\mapsto 0$, since the process is energetically not allowed when Lorentz symmetry is intact. Fourth, since the threshold is proportional to the particle mass, further decay channels into more massive fermion-antifermion pairs open up for rising photon energy.

The total decay width $\Gamma$ of photon decay for the isotropic sector was already obtained in Refs.~\cite{Klinkhamer:2008ky}. Instead, what we will need in the forthcoming analysis are different quantities, which have not been reported in the literature, so far. If the incoming-photon energy exceeds the threshold, the surplus in energy is used to provide a nonzero angle $\theta$ between the 3-momenta of the final-state particles. The latter reads
\begin{equation}
\label{eq:angle-final-fermions}
\cos\theta=\frac{E_f(E_{\gamma}-E_f)+\frac{\tilde{\kappa}_{\mathrm{tr}}}{1-\tilde{\kappa}_{\mathrm{tr}}}E_{\gamma}^2+m_f^2}{\sqrt{[E_f^2-m_f^2][(E_{\gamma}-E_f)^2-m_f^2]}}\,,
\end{equation}
where $E_{\gamma}$ and $E_f$ are the energies of the incoming photon and outgoing fermion, respectively. The energy of the outgoing antifermion is determined by energy-momentum conservation. Furthermore, the partial decay width with respect to the energy of the final fermion can be cast into the form
\begin{align}
\label{eq:partial-decay-width}
\frac{\mathrm{d}\Gamma}{\mathrm{d}E_f}&=\frac{\alpha}{(1+\tilde{\kappa}_{\mathrm{tr}})^2\sqrt{1-\tilde{\kappa}_{\mathrm{tr}}^2}E_{\gamma}^2} \notag \\
&\phantom{{}={}}\times\Big\{(1-\tilde{\kappa}_{\mathrm{tr}})\left[2\tilde{\kappa}_{\mathrm{tr}}E_f(E_{\gamma}-E_f)+(1+\tilde{\kappa}_{\mathrm{tr}})m_f^2\right] \notag \\
&\phantom{{}={}}\hspace{0.6cm}-\tilde{\kappa}_{\mathrm{tr}}E_{\gamma}^2\Big\}\,,
\end{align}
with the fine structure constant $\alpha=e^2/(4\pi)$.  \eqref{eq:angle-final-fermions} and \eqref{eq:partial-decay-width} are new results. Evaluating the final-particle phase space, the electron energy lies within the following interval of a minimum and maximum energy:
\begin{subequations}
\begin{align}
\label{eq:fermion-energy-domain}
E_f&\in [E_{\mathrm{min}},E_{\mathrm{max}}]\,, \\[1ex]
E_{\mathrm{min}}&=\frac{1}{2}(E_{\gamma}-\overline{E})\,,\quad E_{\mathrm{max}}=\frac{1}{2}(E_{\gamma}+\overline{E})\,, \\[1ex]
\overline{E}&\equiv \sqrt{\frac{1+\tilde{\kappa}_{\mathrm{tr}}}{1-\tilde{\kappa}_{\mathrm{tr}}}\left[E_{\gamma}^2+2\left(\frac{1}{\tilde{\kappa}_{\mathrm{tr}}}-1\right)m_f^2\right]}\,.
\label{eq:validity-domain}
\end{align}
\end{subequations}
The angle of \eqref{eq:angle-final-fermions} is well-defined only when $E_f$ lies in the interval set by Eqs.~(\ref{eq:fermion-energy-domain}) -- (\ref{eq:validity-domain}).


As an illustration, we focus on top quarks, assuming a mass $m_{\mathrm{t}}=\unit[172.5]{GeV}$.
We highlight the difference between the kinematics of photon decay to top quark pairs relative to that of SM $\mathrm{pp\rightarrow t\bar{t}}$ processes. 
In this setting, the parameter $\Tilde{\kappa}_{\mathrm{tr}}$ must be replaced by a combination of photon and top-quark coefficients~\cite{Hohensee:2008xz}, $\Tilde{\kappa}_{\mathrm{tr}}-(4/3)c_{\mathrm{t}}^{TT}$. Besides, as $c_{\mathrm{t}}^{TT}$ is an isotropic coefficient in the quark sector where constraints are weak compared to those for $\Tilde{\kappa}_{\mathrm{tr}}$ \cite{Kostelecky:2008ts}, we can neglect $\Tilde{\kappa}_{\mathrm{tr}}$ in the following.
The coefficient $c_{\mathrm{t}}^{TT}$ induces a shift in the cross section for the process $\mathrm{pp}\rightarrow \mathrm{t\bar{t}}$, which would likely be attributed to a QCD effect~\cite{Berger:2015yha,Carle:2019ouy}.
\begin{figure}
    \centering
    \includegraphics[width=1\linewidth]{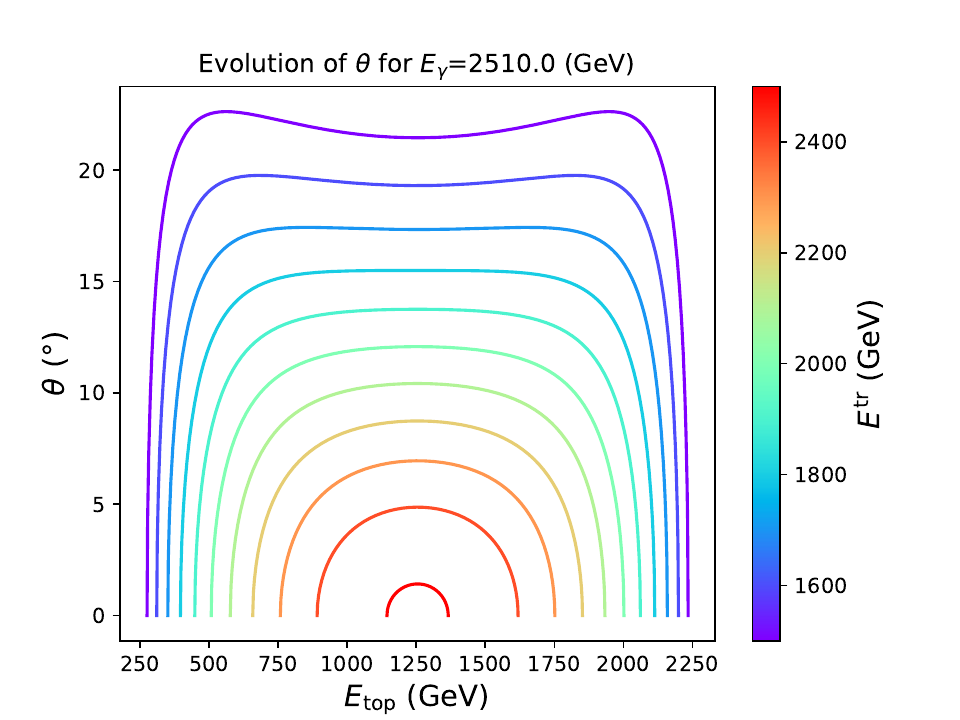}
    \caption{Opening angles between the final top quarks for a photon energy of \unit[2510]{GeV}, at different threshold energies.}
    \label{fig:ThetaTop}
\end{figure}

However, top pairs produced from photon decay would entail a different kinematics, which is illustrated in \figref{fig:ThetaTop} via the opening angle between top quarks for several threshold energies. This finding could be employed to set bounds on $c_{\mathrm{t}}^{TT}$.
It was checked that the azimuthal difference between top quarks shows a distribution of events concentrated at $\Delta\phi \approx 0$ for photons above an energy threshold close to \unit[2.5]{TeV}. Therefore, this leads to $\Delta\phi(l^{+},l^{-}) \approx 0$ in leptonic top-quark decay, as well, which is very different from the predictions in the SM~\cite{CMS:2019nrx, ATLAS:2023gsl}. This region is poorly understood from a theoretical point of view, though, with SM predictions of a toponium bound state~\cite{Fuks:2021xje} yet to be observed. Furthermore, it is not clear if top quarks arising from photon decay at this energy would still decay preferentially to $\mathrm{Wb}$ as predicted in the SM~\cite{Altschul:2020wkw} and a dedicated reconstruction algorithm might be needed. Using this difference of kinematics to constrain LV remains a challenging path in the present state of the theory. However, interpreting LHC data using solely photon disappearance still requires Eqs.~(\ref{eq:relation-kappa-Ethresh}) to (\ref{eq:validity-domain}), as we shall see in the next section.

\section{Search for photon decay to electrons \textit{in vacuo} at the LHC}
\label{sec:section-photon-decay-electrons}

The proton-proton collisions at the LHC are providing a large sample of high $E_{T,\gamma}$ prompt photons (where $E_{T,\gamma}$ is their transverse energy).
In this section, we reinterpret in the context of a possible LV the distribution of the differential cross section $\mathrm{d}\sigma/\mathrm{d}E_{T,\gamma}$ of inclusive prompt photons, which was measured with the ATLAS detector at a center-of-mass energy of $\sqrt{s}=13$ TeV~\cite{ATLAS:2019buk}. This analysis, performed with an integrated luminosity of \unit[36.1]{fb$^{-1}$} at the LHC Run 2, reports photons with the largest $E_{T,\gamma}$ observed at the LHC, binning up to \unit[2.5]{TeV} for photon pseudorapidity $|\eta^{\upgamma}|<0.6$. The latter provides an improvement as compared to a previous result of the CMS experiment, with photons of up to \unit[2]{TeV}~\cite{CMS:2018qao}.

In this section, we make the hypothesis that photons can decay into $\mathrm{e^{+}e^{-}}$.
In the following, models are built for $E_{T,\gamma}$ distributions, based on the hypothesis of LV signals and SM backgrounds.
Events of the process $\mathrm{pp}\rightarrow \upgamma+$jet at tree-level with up to 3 additional partons at leading order (LO) in perturbative QCD (pQCD) and matched with parton showers are simulated with the SHERPA generator v.2.2.15~\cite{Sherpa:2019gpd,Siegert:2016bre}. The events are further reweighted using tables from HEPData~\cite{ATLAS_PromptPhoton_HepData} such that no difference remains in the $E_{T,\gamma}$ distribution between the SHERPA sample used in this article and the sample employed in the ATLAS paper~\cite{ATLAS:2019buk}.
The events are selected using RIVET~\cite{Bierlich:2019rhm}, a framework providing codes for the analysis of parton-shower level Monte-Carlo (MC) events corresponding to publicly available experimental results. We require $E_{T,\gamma} > \unit[125]{GeV}$ within the acceptance region $|\eta^{\upgamma}|<0.6$, with criteria on photon isolation taken from the RIVET routine corresponding to those of a similar analysis performed at $\unit[8]{TeV}$~\cite{ATLAS:2016fta}. Note that the same criteria are employed in the $\unit[13]{TeV}$ analysis~\cite{ATLAS:2019buk}.

For each event passing the selection, the photon with the largest $E_{T,\gamma}$ is selected. Under the hypothesis of a nonzero $\Tilde{\kappa}_{\mathrm{tr}}$, the probability for photons to reach the ATLAS detector is computed as $\mathrm{e}^{-\Gamma x}$, using the total photon decay width $\Gamma$, which follows from integrating \eqref{eq:partial-decay-width}. Moreover, we assume a distance of $\unit[33]{mm}$ between the interaction point and the closest layer of the ATLAS pixel detector~\cite{ATLAS:2010ojh} at LHC Run 2.
As already noted~\cite{Hohensee:2008xz}, the photon decay process is very efficient, since only 0.038\% of the photons do not decay over this distance, above the threshold $E^{\mathrm{th}}=\unit[2]{TeV}$.
If photons decay, distributions based on Eqs.~(\ref{eq:angle-final-fermions}) and (\ref{eq:validity-domain}) are used to generate the fermion and antifermion four-momenta, employing uniform distributions in azimuths.

\begin{figure}
    \centering
    \includegraphics[width=1\linewidth]{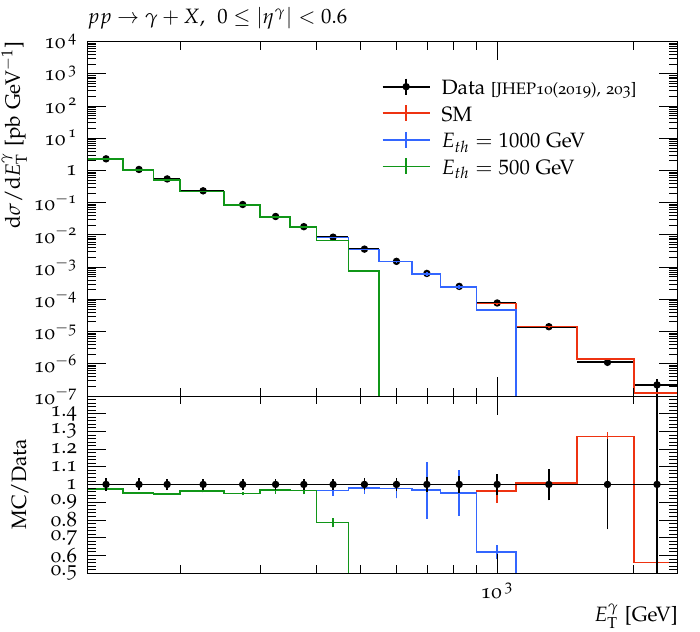}
    \caption{Differential cross section for inclusive prompt photon production measured with the ATLAS detector~\cite{ATLAS:2019buk}, compared with the predictions from the SHERPA generator assuming several energy thresholds for photon decay \textit{in vacuo}.}
    \label{fig:SignalModel}
\end{figure}

Since the electrons arising from photon decay \textit{in vacuo} can be reconstructed within the detector, it must be verified whether or not they would still be identified as photons.
The $\mathrm{e^{+}e^{-}}$ system can be reconstructed as one-leg or two-leg converted photons, or as electrons.
Reference~\cite{ATLAS:2019buk} reports an uncertainty of approximately 1.5\% on photon identification, which we take as an estimate of the probability for an electron to be reconstructed as a photon.
If one electron is reconstructed as a photon, the value of its transverse energy, drawn according to the probability density $\mathrm{d}\Gamma/\mathrm{d}E_f$, is included in the distribution of $E_{T,\gamma}$. If two electrons are reconstructed as a photon, the largest transverse energy is included in the distribution of $E_{T,\gamma}$.
Among the photons within a $E_{T,\gamma}$ bin, a fraction of about 3\% would be recovered from the decay into $\mathrm{e^{+}e^{-}}$ as reconstructed photons with a lower $E_{T,\gamma}$ value.
The procedure described above provides models for the  $E_{T,\gamma}$ distribution of SM prompt photon production (treated as a background), and of a SM contribution with LV (treated as a signal) under any hypothesis for the value of $\Tilde{\kappa}_{\mathrm{tr}}$. These can be compared to the spectrum of $E_{T,\gamma}$ measured with the ATLAS detector. Predictions are shown in \figref{fig:SignalModel} for several values of the threshold energy.

The statistical treatment employed to extract bounds on $\Tilde{\kappa}_{\mathrm{tr}}$ uses the modified frequentist $CL_s$ method~\cite{Read:2002hq}, based on the likelihood ratio of the SM+LV hypothesis against the SM-only hypothesis, assuming the number of events are Poisson-distributed in each bin.
The central value for the number of events $N_i$ is computed from the differential cross sections $\mathrm{d}\sigma_i/\mathrm{d}E_{T,\gamma}$ in each bin $i$ as:
\begin{equation}
\label{eq:number-of-events-from-diff-xs}
N_i = \frac{\mathrm{d}\sigma_i}{\mathrm{d}E_{T,\gamma}} \cdot \Delta E_{T,\gamma,i} \cdot \epsilon \cdot L\,,
\end{equation}
where $\Delta E_{T,\gamma}$ is the bin width, $\epsilon$ the efficiency of photon reconstruction and identification, and $L$ is the integrated luminosity.
The systematic uncertainties in the ATLAS measurement, which impact the predictions for the models, are from multiple sources: background subtraction, unfolding of the detector effect, pileup, trigger and photon selection efficiencies, photon energy scale, and resolution.
Other uncertainties considered are the MC statistical uncertainty in the SHERPA sample, as well as the theory uncertainties quoted for the ATLAS sample, arising from QCD renormalization and factorization scale, the parton distribution functions, strong coupling constant $\alpha_S$, and the parton shower.
The value of the systematic uncertainties are added in quadrature in each bin, and treated conservatively as a separate Gaussian nuisance parameter for each bin of the $E_{T,\gamma}$ distribution. The total systematic uncertainty ranges from 2.3\% at low values of $E_{T,\gamma}$ to 7.6\% at high values.

\begin{figure}
    \centering
    \includegraphics[width=1\linewidth]{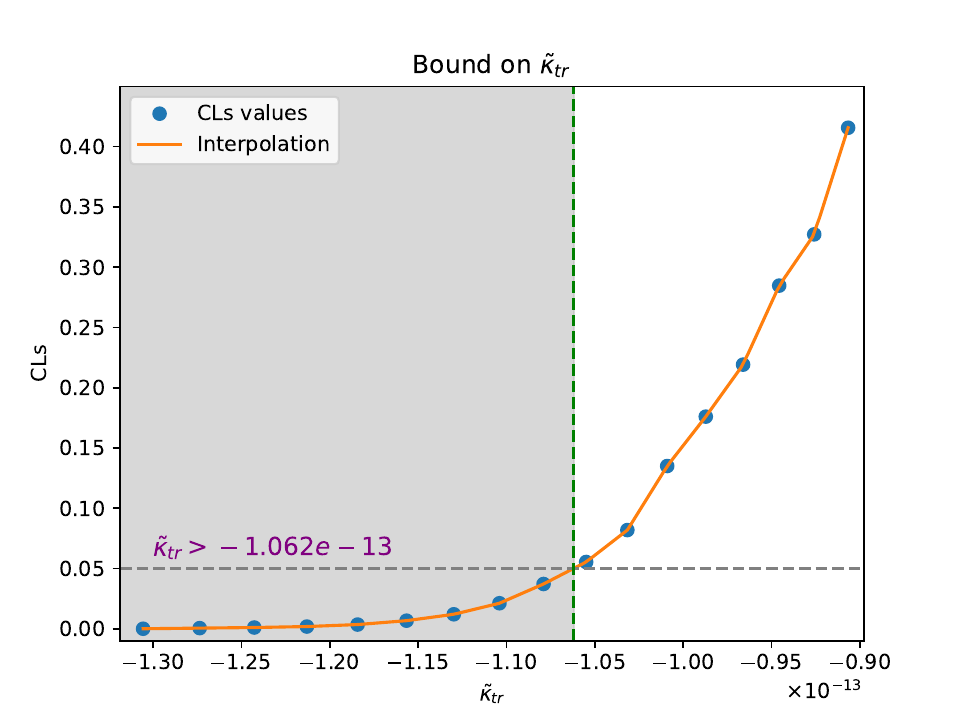}
    \caption{Bound on $\Tilde{\kappa}_{\mathrm{tr}}$ as a function of the $CL_s$ value. A bound with a 95\% confidence level is defined by the region $CL_s<0.05$.}
    \label{fig:KappaCLs}
\end{figure}

The achieved bound on $\Tilde{\kappa}_{\mathrm{tr}}$ as a function of the $CL_s$ criterion is shown in Fig.~\ref{fig:KappaCLs}. The conventional criterion of $CL_s<0.05$ is employed, leading to the following limit at 95\% confidence level on the allowed region:
\begin{equation}
\label{eq:main-result}
\Tilde{\kappa}_{\mathrm{tr}} > -1.06 \times 10^{-13}\,.
\end{equation}
This corresponds to a threshold energy of $E^{\mathrm{th}}=\unit[2.218]{TeV}$.
The bound quoted above represents a large improvement, of a factor of $\approx 55$, over the previous bound $\Tilde{\kappa}_{\mathrm{tr}} > -5.8 \times 10^{-12}$~\cite{Hohensee:2008xz}. The latter was set by reinterpreting D0 prompt photon data~\cite{D0:2008chx} where the last $E_{T,\gamma}$ bin was measured up to $\unit[340]{GeV}$.
The order of magnitude of the improvement can be understood, since the LV/SM separation is the largest in the last bin, reporting photons with the largest energy, and recalling that $\Tilde{\kappa}_{\mathrm{tr}} \propto (1/E^{\mathrm{th}})^2$ according to \eqref{eq:relation-kappa-Ethresh}.
As a matter of fact, assuming conservatively that the   the largest energy observed is set by the left-hand side boundary of the last measured bin, i.e., $\unit[2]{TeV}$, and that it corresponds to $\eta^{\upgamma}=0$, fixes the minimum threshold to $E^{\mathrm{th}}\approx \unit[2]{TeV}$.

By assuming that all of the higher-energy photons decay immediately, this in turn provides a qualitative bound $\Tilde{\kappa}_{\mathrm{tr}} > -1.3 \times 10^{-13}$.
It should be noted that this method leads to a qualitative estimate only, since it does not include systematic uncertainties, and neither considers the probability of photons to survive up to the detector nor the probability of the $\mathrm{e^{+}e^{-}}$ system to be reconstructed as a photon.
The improvement of \eqref{eq:main-result} by nearly 20\% over this simple estimate arises from the statistical method employed, which makes full use of the statistical power of the last $E_{T,\gamma}$ bin.  
It is instructive to check the result without any systematic uncertainties, in which case the bound achieved would be $\Tilde{\kappa}_{\mathrm{tr}} > -1.045 \times 10^{-13}$. The result in \eqref{eq:main-result} is dominated by the statistical uncertainties in the data.

An upper bound $\Tilde{\kappa}_{\mathrm{tr}} < 1.2 \times 10^{-11}$ was set from the LEP beam energy stability~\cite{Hohensee:2008xz}, being interpreted as showing no evidence for vacuum Cherenkov radiation from electrons. Such a limit cannot be improved by referring to the LHC beam stability, since the LHC collides protons. The SuperKEKB facility~\cite{Ohnishi:2013fma} is nowadays colliding $\mathrm{e}^{+}$ and $\mathrm{e}^{-}$, but does so at a lower center-of-mass energy than at LEP, which still provides the best results.
Future particle colliders like the FCC-ee would collide electrons of $\unit[183]{GeV}$~\cite{FCC:2018evy} providing upper bounds improved by a factor of 3. The FCC-hh~\cite{FCC:2018vvp} would collide $\unit[100]{TeV}$ protons; assuming that prompt photons of $\approx\unit[20]{TeV}$ would be produced, the lower bound would improve by 2 orders of magnitude over the results presented in this paper.

\section{Conclusions}
\label{sec:conclusions}

This paper provides a new evaluation of the kinematics of photon decay into fermions \textit{in vacuo} under an isotropic violation of Lorentz invariance.
These results were used to reinterpret a measurement of inclusive prompt photon production at the LHC Run 2, with photons observed up to a transverse energy of \unit[2.5]{TeV}.
The case of photon decay to top quarks illustrates the change in kinematics relative to the SM predictions.
A new bound on the isotropic SME coefficient $\Tilde{\kappa}_{\mathrm{tr}}$ is set from LHC data, arising from the absence of an observation of photon decay to electrons \textit{in vacuo}. The limit yields $\Tilde{\kappa}_{\mathrm{tr}} > -1.06 \times 10^{-13}$ at 95\% confidence level, a result that is approximately 55 times more stringent than a previous bound extracted from Tevatron data. The calculations for the kinematics of photon decay could be used in the future to bound LV coefficients from the appearance of fermion pairs.

\section*{Acknowledgments}

M.S. greatly acknowledges support via the grants FAPEMA Universal 00830/19 and CNPq Produtividade 310076/2021-8. Furthermore, M.S. is indebted to CAPES/Finance Code 001.

\appendix


\end{document}